\documentclass[prl,aps,twocolumn,amsfonts,showpacs,superscriptaddress]{revtex4}
\usepackage{epstopdf}
\usepackage{epsfig}
\usepackage{psfrag}
\usepackage{color}
\usepackage{amsmath}
\usepackage{amssymb}
\usepackage{graphicx}

\def\be{\begin{equation}}
\def\ee{\end{equation}}
\def\bea{\begin{eqnarray}}
\def\eea{\end{eqnarray}}

\begin{document}

\title{Magnetothermoelectric Response at a Superfluid--Mott Insulator Transition}
\author{M. J. Bhaseen}
\affiliation{Rudolf Peierls Centre for Theoretical Physics, 1 Keble Road, Oxford, OX1 3NP, UK.}
\author{A. G. Green}
\affiliation{School of Physics and Astronomy, University of St Andrews, North Haugh, St Andrews, Fife, KY16 9XP, UK.} 
\author{S. L. Sondhi}
\affiliation{Department of Physics, Princeton University, Princeton, NJ 08544, USA.}
\date{\today}

\begin{abstract}
We investigate the finite temperature magnetothermoelectric response 
in the vicinity 
of a superfluid--Mott insulator quantum phase transition.  
We focus on the particle-hole symmetric transitions of 
the Bose--Hubbard model, and combine 
Lorentz invariance arguments with quantum Boltzmann calculations. By means of an epsilon 
expansion, we find that a non-vanishing thermoelectric tensor and  
a finite thermal conductivity are supported in this quantum critical regime. We comment on the singular Nernst effect in
this problem.
\end{abstract}

\maketitle


Since the discovery of high temperature superconductivity, quantum 
phase transitions between Mott insulators and superfluids/superconductors have
been the focus of intense scrutiny. More recently, remarkable advances
in the control of cold atomic gases 
have allowed the observation of
such phase transitions in systems of bosonic atoms \cite{Greiner:SI}. Such transitions separate two of the most fascinating highly correlated 
states 
of matter: the superfluid, which reveals the dramatic consequences of 
quantum coherence on the macroscopic scale, and the Mott insulator, which affirms the impact of 
strong interactions. In particular, a continuous quantum phase transition between the two 
clearly requires a combined treatment of strong quantum fluctuations and interactions. 
Understanding this interplay is useful not only to the particular systems at hand, but to a wide variety of condensed matter problems \cite{Sachdev:QPT}. 

Our general objective in this present work is to advance the theory of 
continuous superfluid--Mott insulator (S-MI) transitions in bosonic systems. 
In particular, we wish to focus on their quantum critical regimes, and to 
examine the response to applied electric and magnetic fields, and 
temperature gradients. We are motivated by measurements of such magnetothermoelectric response 
by Ong and co-workers  \cite{Xu:Vortex,Wang:Dia}. These have
yielded crucial insights into the importance of superconducting fluctuations
in the cuprates \cite{Ussishkin:Gaussian}, although a direct 
correspondence 
with our present bosonic system is clearly lacking.
Our results apply most simply to charged bosons, 
although there are plausible suggestions to generate 
pseudo magnetic fields in neutral atomic systems; 
see e.g. \cite{Sorensen:FQH}.

Specifically,
we focus on the most straightforward example of a S-MI quantum phase transition --- the
particle-hole symmetric version occurring at the tips of the Mott lobes in the
ubiquitous Bose--Hubbard model \cite{Sachdev:QPT} --- and the quantum critical
regime of this transition. 
This model arises in numerous contexts including $^4{\rm He}$ layers, 
thin film superconductors 
and Josephson junction arrays. More recently, this system has also been realized in ground breaking experiments on cold atomic gases \cite{Greiner:SI}, where it is possible to tune through the S-MI transition using lasers. 
This approach confirms the predicted phase diagram \cite{Fisher:Bosonloc} and offers a unique 
handle on these strongly correlated systems.

In view of this broad spectrum of applications, the Bose--Hubbard model has received considerable theoretical attention, see for example \cite{Fisher:Presence,Fisher:Bosonloc,Cha:Universal,Damle:Nonzero,Green:Nonlinear,Green:current}
and references therein. 
Here we shall incorporate the significant effect of an applied magnetic field on the quantum
critical regime of the particle-hole symmetric transition.
This paper is structured as follows: We begin with a short introduction to the 
Bose--Hubbard model and its quantum phase transitions. We recall the quantum Boltzmann equation (QBE) and the 
epsilon expansion used to access the finite temperature critical dynamics \cite{Damle:Nonzero}. We briefly 
discuss the response to a single external field, 
and then move on to the situation in crossed electric and magnetic fields. We combine Lorentz invariance arguments with Boltzmann calculations to provide the fundamental response coefficients. We then examine the complementary situation in a crossed thermal gradient and a magnetic field. We conclude with implications for possible Nernst measurements, a brief summary, and directions for further research.

{\em Model} --- The Bose--Hubbard model describes bosons hopping on a lattice with amplitude $t$, and interacting via the short range repulsive interaction $U$:\begin{equation}
H=-t\sum_{\langle ij\rangle}(b_i^\dagger b_j+b_j^\dagger b_i)-\mu \sum_i n_i+\frac{U}{2}\sum_in_i(n_i-1).
\label{bh}
\end{equation}
The Bose creation and annihilation operators satisfy the usual commutation relations, $[b_i,b_j^\dagger]=\delta_{ij}$, $n_i=b_i^\dagger b_i$ is the number of bosons at site $i$, and $\mu$ is the chemical potential.
 In the context of a Josephson array or superconductor, the bosons represent Cooper pairs of charge $Q=2e$, tunnelling between superconducting regions. In general, one may also include the effects of disorder and long range interactions into such a model, but here we shall concentrate on the simplest case (\ref{bh}). The phase diagram of the Bose--Hubbard model exhibits both superfluid and Mott insulating regions \cite{Fisher:Bosonloc}, the latter occurring for strong enough repulsive interactions. In particular, as a function of the chemical potential, $\mu$, this model exhibits a series of Mott insulating ``lobes'' where the density of bosons is pinned to successive integers. At the tip of these lobes, the energy cost to producing particle--hole excitations vanishes and the model is particle--hole symmetric. In the vicinity of these points, the S-MI transition is described by the continuum action of a complex scalar field $\Phi$ \cite{Fisher:Bosonloc},
\begin{equation}
S=\int d^Dx\,|\partial_\mu\Phi|^2-m^2|\Phi|^2-\frac{u_0}{3}|\Phi|^4,
\label{SLG}
\end{equation}
where $D=d+1$, and $d$ is the number of spatial dimensions. This is nothing but a quantum Landau--Ginzburg theory for the superconducting order parameter, $\Phi$. The bare couplings $m$ and $u_0$ may be related to the microscopic parameters of the Bose--Hubbard model, and we refer the reader to \cite{Sachdev:QPT} for more details. Away from these particle--hole symmetric points, the effective action picks up an additional term linear in the time derivative. Correspondingly, the dynamical exponent changes from $z=1$ to $z=2$ \cite{Sachdev:QPT}. For the purposes of this paper we will focus on the case with $z=1$. 

{\em Quantum Boltzmann Equation} --- In the absence of interactions, the model (\ref{SLG}) is the celebrated Klein--Gordon theory. The two modes of opposite 
charge correspond to density fluctuations, or particle--hole excitations, of the Mott insulator. 
Performing this mode expansion on the full model (\ref{SLG}) allows one to develop a Boltzmann approach to transport which is particularly simple at non-zero temperatures and small applied fields \cite{Damle:Nonzero}. The resulting quantum Boltzmann equation (QBE) is a nonlinear integro-differential equation for the momentum space distribution functions, $f_\pm({\bf k},t)$, of particle and hole excitations 
\begin{equation}
\frac{{\partial f}_{\pm}}{\partial t}\pm Q\left({{\bf E}+{\bf v}_{\bf k}\times{\bf B}}\right).\frac{\partial f_\pm}{\partial{\bf k}}={\rm I}_\pm[f_+,f_-],
\label{QBE}
\end{equation}
where ${\bf v}_{\bf k}\equiv \partial {\varepsilon}_{\bf k}/\partial {\bf k}$ and $\varepsilon_{\bf k}\equiv \hbar\omega_k=\sqrt{{\bf k}^2+m^2}$. For simplicity we consider a spatially homogeneous system in uniform external fields. The collision term represents scattering between these excitations, and incorporates the nonlinear interaction of the Landau--Ginzburg action (\ref{SLG}) in a systematic epsilon expansion \cite{Damle:Nonzero}: 
\begin{eqnarray}
{\rm I}_\pm & = & -\frac{2u_0^2}{9}\int \prod_{i=1}^3 \frac{d^d{\bf k}_i}{(2\pi)^d\,2\varepsilon_{k_i}}\left(\frac{{\mathcal F}_\pm^{\rm out}-{\mathcal F}_\pm^{\rm in}}{2{\varepsilon}_k}\right)\times  \nonumber \\
& & \hspace{-1cm} (2\pi)^{d+1}\delta({\bf k}+{\bf k}_1-{\bf k}_2-{\bf k}_3)\delta(\varepsilon+\varepsilon_1-
\varepsilon_2-\varepsilon_3),
\label{col}
\end{eqnarray}
where scattering out of state ${\bf k}$ is given by
\begin{eqnarray}
{\mathcal F}_\pm^{\rm out} & = & 2f_\pm({\bf k})f_\mp({\bf k}_1)[1+{f}_\pm({\bf k}_2)][1+ {f}_\mp({\bf k}_3)] \nonumber \\
& & + f_\pm({\bf k})f_\pm({\bf k}_1)[1+{f}_\pm({\bf k}_2)][1+{f}_\pm({\bf k}_3)],\nonumber \end{eqnarray}
and we have suppressed the explicit time dependence of the distribution functions. Scattering 
into state ${\bf k}$ follows by interchanging $f_\pm$ and $1+f_\pm$. At the Wilson--Fisher fixed point, where the renormalized mass vanishes, the bare couplings must be tuned to the values \cite{Damle:Nonzero}
\begin{equation}
m^2=\frac{4\pi^2T^2\epsilon}{15}, \quad u_0=\frac{24\pi^2\epsilon}{5}.
\end{equation}
The structure of the collision term (\ref{col}) may also be seen using Fermi's Golden rule, where the $1+f$ factors remind 
us that we are dealing with a system of bosons. In this representation the electric current takes the form 
\begin{eqnarray}
{\bf J}_e=Q \int \frac{d^d k}{(2\pi\hbar)^d}\,{\bf v}_{\bf k}\,[f_+({\bf k},t)-f_-({\bf k},t)]
\label{bolel}
\end{eqnarray}
and the heat current is given by
\begin{equation}
{\bf J}_h=\int \frac{d^d k}{(2\pi\hbar)^d}\,{\bf v}_{\bf k}\varepsilon_{\bf k}\,[f_+({\bf k},t)+f_-({\bf k},t)].
\label{bolheat}
\end{equation}
In writing this last expression we have used the fact that within linear response, the heat current coincides with the energy current. The transport coefficients of interest are defined by the relationships
\begin{equation}
\begin{pmatrix} {\bf J}_e^{\rm tr} \\ {\bf J}_h^{\rm tr} \end{pmatrix} =\begin{pmatrix}\sigma & \alpha \\ \tilde\alpha & \kappa
\end{pmatrix}\begin{pmatrix}{\bf E}\\ -\nabla T\end{pmatrix}.
\label{transcoeffs}
\end{equation}
In order to work below $d=3$ in the epsilon expansion, we will restrict ourselves to working
with ${\bf E}$ and $\nabla T$ orthogonal to $\bf B$, and to transport 
coefficients in the ``xy'' plane defined by $\bf E$ and $\bf E \times B$.  
Particle-hole symmetry places constraints on these 
remaining coefficients: the longitudinal component of $\alpha$ and the 
transverse (Hall) components of $\sigma$ and $\kappa$ must vanish. 
Finally, we note that here, as in the standard Boltzmann treatment of 
magnetothermoelectric transport, we do not explicitly
consider magnetization currents induced by the applied magnetic 
field \cite{Cooper:Thermo,Ussishkin:Gaussian}.
We will discuss the implicit cancellations behind 
this elsewhere \cite{BGS:inprog}.

{\em Separate response} --- Before we turn our attention to the general problem in combined fields, 
let us first discuss what happens when each of the fields, ${\bf E}$, $\nabla T$, and ${\bf B}$, is taken separately. The application of an electric field at finite temperatures leads to a linear response conductivity, 
$\sigma_{xx}$, calculated in \cite{Damle:Nonzero}. While the model thus 
supports current relaxation,
it does not support energy current relaxation; it is known that its 
response to an applied
temperature gradient---the thermal conductivity $\kappa_{xx}$--- is infinite \cite{Vojta:RG2}. 
The inclusion of the highly irrelevant 
umklapp scattering, reminiscent of the one-dimensional problems studied in \cite{Shimshoni:Thermal}, will ultimately render this quantity finite, though anomalously large as $T \rightarrow 0$.
Finally, let us now turn to the response to a magnetic field taken alone. In the quantum critical region,
and in the absence of an electric field and temperature gradient, the only 
relevant energy scales are the temperature and the magnetic field. On general grounds we expect 
the free energy density to scale as 
\begin{equation}
{\mathcal F}(T,B)=T^{1+d/z}\,f_1\left(\frac{B}{T^{2/z}}\right),
\end{equation}
where $f_1$ is some scaling function. Here we have used the fact that the correlation length, $\xi$, diverges with the correlation time, $\xi_t\sim 1/T$, according to $\xi\sim(\xi_t)^{1/z}$. The prefactor is thus an energy density. In addition we have used the fact that the vector potential, $|{\bf A}|\sim 1/L$, as may be seen from its gauge transformations, and so $|{\bf B}|\sim 1/L^2\sim T^{2/z}$. It follows that the linear response magnetization for example, scales as
\begin{equation}
M=-\frac{\partial {\mathcal F}}{\partial B}\sim T^{1+(d-4)/z}B.
\end{equation}
This is consistent with a finite temperature, diagrammatic Kubo calculation of the 
magnetic susceptibility of a charged scalar field, with $m\sim T$ and $z=1$ \cite{BGS:inprog}.

{\em Crossed ${\bf E}$ and ${\bf B}$ fields} --- The motion of a single relativistic charged particle in crossed electric and magnetic fields has two distinct regimes of behavior \cite{Jackson:Classical}. In the regime where $|{\bf E}|<c|{\bf B}|$ one may always find a frame moving with the velocity
\begin{equation}
{\bf v}_{\rm D}=\frac{{\bf E}\times{\bf B}}{|{\bf B}|^2},
\label{dv}
\end{equation}
where the electric field vanishes. In this moving frame the particle executes cyclotron orbits. Boosting back to the lab frame the trajectories are helical, with a well defined, charge independent, transverse drift velocity (\ref{dv}). In the regime $|{\bf E}|>c|{\bf B}|$ it is possible to find a frame where the magnetic field vanishes, but it is no longer possible to make the electric field vanish. In consequence, in the absence of any dissipative processes,  the particle is continually accelerated by the electric field. In the lab frame the trajectories are hyperbolic. Since we are interested in linear response in a background magnetic field, we shall focus primarily on the drift regime; we
comment on the hyperbolic regime later. The speed of light at issue here is the bosonic mode velocity
suppressed in Eqn. (\ref{SLG}).

{{\em Drift regime} $|{\bf E}|<c|{\bf B}|$} --- So far we have used the Lorentz transformations to note some properties in the absence of the collision term. However, the  Lorentz structure also enables us to make 
progress with the full QBE (\ref{QBE}). As discussed above, the electric field vanishes in a frame moving with velocity ${\bf v}_{\rm D}$. Since a magnetic field does not affect the energy of a particle it follows that a thermal distribution holds in this frame. Using the Lorentz transformation for energy this suggests that
\begin{equation}
f_\pm({\bf k})=f_0(\varepsilon_k^\prime)=f_0\left(\frac{\varepsilon_k-{\bf v}_{\rm D}.{\bf k}}{\sqrt{1-v_{\rm D}^2/c^2}}\right),
\label{boostdist}
\end{equation}
is a solution of the full QBE (\ref{QBE}), {\em including the collision 
term.} This remarkable fact may be verified by explicit substitution. In particular, we may Taylor expand (\ref{boostdist}) to linear order in ${\bf v}_{\rm D}$, or equivalently the electric field. Substituting this into (\ref{bolheat}) 
yields a non-vanishing transverse heat current, and a corresponding thermoelectric tensor
\begin{equation}
\alpha_{xy}=\frac{2c^2}{dBT}\int\frac{d^dk}{(2\pi\hbar)^d}\,k^2
\left(-\frac{\partial f_0}{\partial \varepsilon_k}\right).
\label{alphad}
\end{equation}
This result also follows from entropy drift \cite{BGS:inprog}. Note that, we may also make contact with (\ref{boostdist}) from a direct linearization of the QBE. Substituting the decomposition 
\begin{equation}
f_\pm({\bf k})=f_0(\varepsilon_k)\pm Q{\bf k}.{\bf E}\,\psi(k)+{\bf k}.({\bf E}\times{\bf B})\,\psi_\perp(k)
\end{equation}
into equation (\ref{QBE}) and linearizing in the electric field, yields $\psi(k)=0$, and an expression for $\psi_\perp(k)$ consistent with the Taylor expansion of (\ref{boostdist}).
Having established the validity of (\ref{alphad}) as a legitimate result for the full QBE, we may complete the
final steps of the epsilon expansion. Following Damle and Sachdev \cite{Damle:Nonzero}, to lowest order we compute the numerical prefactor in $d=3$ and for $m=0$:
\begin{equation}
\alpha_{xy}=\frac{4\pi^2}{45}\frac{k_B}{B}
\left(\frac{k_BT}{\hbar c}\right)^{3-\epsilon}.
\label{alphaeps}
\end{equation}
The form of this result is consistent with scaling arguments in the vicinity 
of a QCP \cite{BGS:inprog}
\begin{equation}
\alpha_{xy}(T,B)\sim T^{(d-2)/z}\,f_2\left(\frac{B}{T^{2/z}}\right)\sim 
\frac{T^d}{B},
\end{equation}
where $f_2$ is another scaling function. Note that the dependence on $1/B$, which stems from the drift 
velocity (\ref{dv}), is analogous to that of the clean, single carrier classical Hall conductivity. It is important to recall however, that $\sigma_{xy}=0$, in our particle-hole symmetric, two species case. Likewise, $\sigma_{xx}=0$ in this regime, owing to the purely transverse nature of the distribution function (\ref{boostdist}). Having completed our analysis in the drift regime, 
let us discuss the transport properties in the presence 
of a thermal gradient and a magnetic field.

 {\em Crossed $\nabla T$ and ${\bf B}$ fields}  --- In keeping with our previous discussion, let us consider $|\nabla T|\ll |{\bf B}|$. We may introduce a temperature gradient into the QBE by allowing the temperature variable to be a function of position \cite{Ziman:Principles}. The generic QBE, with ${\bf E}=0$, reads
\begin{equation}
\frac{\partial f_\pm}{\partial t}+
{\bf v}_k.\frac{\partial f_\pm}{\partial{\bf x}}\pm
Q({\bf v}_{\bf k}\times{\bf B}).\frac{\partial f_\pm}{\partial {\bf k}}
={\rm I}_\pm[f_+,f_-].
\label{QBE2}
\end{equation}
In the absence of any material inhomogeneity we may assume that any spatial variation is due to the imposed temperature gradient:
\begin{equation}
\frac{\partial f_\pm}{\partial {\bf x}}=
\nabla T\left(\frac{\partial f_\pm}{\partial T}\right)=
\nabla T\left(
-\frac{\varepsilon_k}{T}\frac{\partial f_0}{\partial \varepsilon_k}\right).
\end{equation}
In the last step we have used the fact that we are interested in linear response in $\nabla T$. To solve the resulting linearized equation we parametrize
\begin{equation}
f_\pm({\bf k})=f_0(\varepsilon_k)+{\bf k}.{\bf U}\,\psi(k)\pm Q{\bf k}.({\bf U}\times {\bf B})\,\psi_\perp(k).
\label{betexp}
\end{equation}
To lowest order in the epsilon expansion we may drop the collision terms to obtain
\begin{equation}
\psi(k)=0,\quad \psi_\perp(k)=\frac{\varepsilon_k}{Q^2|{\bf B}|^2}\left(-\frac{\partial f_0}{\partial \varepsilon_k}\right).
\label{tempperp}
\end{equation}
This reproduces our previous result (\ref{alphad}) as expected from Onsager reciprocity. 

Proceeding to ${\mathcal O}(\epsilon^2)$ we find a non-vanishing {\em longitudinal} displacement
\begin{equation}
\psi(k)  =  \epsilon^2\left(\frac{\varepsilon_k}{\hbar c}\right)
\int_0^\infty dk_1\left[\psi_\perp(k){\rm F}_1 +\psi_\perp(k_1){\rm F}_2 \right],
\label{psieps}
\end{equation}
where $\psi_\perp(k)$ is given by our zeroth order result (\ref{tempperp}), and ${\rm F}_1(k,k_1)$ and ${\rm F}_2(k,k_1)$ are the  kernels employed by Damle and Sachdev \cite{Damle:Nonzero}. 
Introducing the dimensionless variable, ${\bar k}=ck/k_BT$, the longitudinal displacement may be written
\begin{equation}
\psi(k)=\frac{\epsilon^2}{\hbar}\left(\frac{k_BT}{Q B c}\right)^2 G({\bar k}),
\label{glongi}
\end{equation}
where
\begin{equation}
G(\bar k) \equiv  {\bar k}\int_0^\infty d{\bar k}_1
\left[{\rm P}({\bar k})\Phi_1({\bar k},{\bar k}_1)+\Phi_2({\bar k},{\bar k}_1)
{\rm P}({\bar k}_1)\right].
\label{gfunc}
\end{equation}
Here ${\rm P}({\bar k})={\bar k}e^{\bar k}/(e^{\bar k}-1)^2$ is the massless limit of $\psi_\perp(k)$ and $\Phi_{1,2}$ are the rescaled kernels \cite{Damle:Nonzero}. Upon integration over momentum space, the longitudinal displacement leads to a {\em finite} thermal conductivity 
\begin{equation}
\kappa=g\,\epsilon^2(k_Bc)\,{l}_B^4\,\lambda_T^{-(d+3)},
\label{kaplengths}
\end{equation}
where $l_B\equiv (\hbar/QB)^{1/2}$ and $\lambda_T\equiv \hbar c/k_BT$ are the magnetic length, and a suitable thermal wavelength respectively. To lowest order in the epsilon expansion the numerical coefficient is given by 
\begin{equation}
g=\frac{1}{3\pi^2}\int_0^\infty d{\bar k}\,{\bar k}^{4}\,G({\bar k})\approx 
5.55.
\end{equation}
This equation mirrors (3.33) of Damle and Sachdev \cite{Damle:Nonzero}.
The extra factor of momentum arises because we are considering {\em
  heat} transport as opposed to electrical transport. In addition, the
scaling function $G({\bar k})$ is determined by the direct integration
given in equation (\ref{gfunc}). It is {\em not} obtained by
solving a linear integral equation such as (3.31) of \cite{Damle:Nonzero}. 

{\em Beyond the linear Boltzmann regime} --- Above we have considered linear transport response 
in $\bf E$ and $\nabla T$ with a finite $B \ll T$ already present. This is what is naturally
described by the linearized Boltzmann equation. Evidently, this is not all the behavior
that is possible in the quantum critical regime: with $T \gg E,B, \nabla T$, it is still possible
to have other ratios for the three applied fields. We have previously noted that
for $E>cB$ single particle orbits are fundamentally different. One can show
more generally that the drift solution fails to exist in this domain, even in the presence
of the interactions contained in the fixed point theory (\ref{SLG}). It seems clear that
the fixed point theory should yield a finite $\sigma_{xx}$, and an infinite $\alpha_{xy}$
in this regime, but a formal demonstration needs to be constructed. The 
situation with
an applied $\nabla T$ is similar, but in need of more work. Specifically, the boundary 
between the $\nabla T \ll B$ regime with a finite $\kappa_{xx}$, and the $\nabla T \gg B$ 
regime which is expected to exhibit an infinite $\kappa_{xx}$, although finite 
$\alpha_{xy}$, cannot be located with
precision by our methods. We hope to report progress on these remaining issues elsewhere \cite{BGS:inprog}.

{\em Concluding remarks} --- 
We have examined the magnetothermoelectric response at the 
S-MI transitions of the Bose--Hubbard model. In linear transport in an
applied magnetic field, 
we obtain a finite thermoelectric tensor and thermal conductivity. 
It is worth noting that in contrast to the $B=0$ 
electrical conductivity \cite{Damle:Nonzero} 
these coefficients are regular in the epsilon expansion. Finally, we comment 
on a Nernst measurement in such a system in which we measure a transverse electric
field induced by a thermal gradient in an open electrical circuit. For the measurement
to be well defined, we will need to invoke irrelevant umklapp processes to regularize 
quantities that are infinite in the fixed point theory. This done, we find that
in the scaling limit, an infinitesimal $\nabla T$ induces a finite Nernst electric field of strength $B/c$.

{\em Acknowledgements} --- We are extremely grateful to D. Basko, J.-S. Caux, 
J. Chalker, K. Damle, F. Essler, C. Hooley, A. Lamacraft, S. Sachdev, 
V. Oganesyan, and A. Tsvelik for valuable input at various stages. 
We are especially indebted to D. Huse for 
insightful comments on the Lorentz transformations for $E>cB$. 
This work was supported by NSF grant DMR0213706, EPSRC, and The Royal Society. 
MJB would also like to thank BNL and St Andrews for hospitality during part of this work.


\end{document}